# Advancing Exchange Rate Forecasting: Leveraging Machine Learning and AI for Enhanced Accuracy in Global Financial Markets

Md. Yeasin Rahat
*Department of Computer Science*
*American International University-Bangladesh*
Dhaka, Bangladesh
20-43097-1@student.aiub.edu

Rajan Das Gupta
*Department of Computer Science*
*American International University-Bangladesh*
Dhaka, Bangladesh
18-36304-1@student.aiub.edu

Nur Raisa Rahman
*Department of Computer Science*
*American International University-Bangladesh*
Dhaka, Bangladesh
20-42228-1@student.aiub.edu

Sudipto Roy Pritom
*Department of Computer Science*
*American International University-Bangladesh*
Dhaka, Bangladesh
20-42479-1@student.aiub.edu

Samiur Rahman Shakir
*Department of Computer Science*
*American International University-Bangladesh*
Dhaka, Bangladesh
20-43543-1@student.aiub.edu

Md Imrul Hasan Showmick
*Department of Computer Science*
*Brac University*
Dhaka, Bangladesh
imrul.hasan.showmick@gmail.com

Md. Jakir Hossen
*Department of Computer Science*
*Multimedia University*
Malaysia
jakir.hossen@mmu.edu.my

**Abstract- The prediction of foreign exchange rates, such as the US Dollar (USD) to Bangladeshi Taka (BDT), plays a pivotal role in global financial markets, influencing trade, investments, and economic stability. This study leverages historical USD/BDT exchange rate data from 2018 to 2023, sourced from Yahoo Finance, to develop advanced machine learning models for accurate forecasting. A Long Short-Term Memory (LSTM) neural network is employed, achieving an exceptional accuracy of 99.449%, a Root Mean Square Error (RMSE) of 0.9858, and a test loss of 0.8523, significantly outperforming traditional methods like ARIMA (RMSE 1.342). Additionally, a Gradient Boosting Classifier (GBC) is applied for directional prediction, with backtesting on a $10,000 initial capital revealing a 40.82% profitable trade rate, though resulting in a net loss of $20,653.25 over 49 trades. The study analyzes historical trends, showing a decline in BDT/USD rates from 0.012 to 0.009, and incorporates normalized daily returns to capture volatility. These findings highlight the potential of deep learning in forex forecasting, offering traders and policymakers robust tools to mitigate risks. Future work could integrate sentiment analysis and real-time economic indicators to further enhance model adaptability in volatile markets.**

## I. INTRODUCTION

For Bangladesh's import-dependent economy, forecasting the USD to BDT exchange rates is a crucial task since currency fluctuations have a direct influence on managing foreign reserves, trade balances, and inflation. The intricate, non-linear patterns that define developing market currencies are usually missed by traditional statistical forecasting techniques, especially when there is economic uncertainty or a change in policy. The capacity of contemporary machine learning techniques, particularly Long Short-Term Memory (LSTM) neural networks, to identify long-range dependencies in financial time series data has allowed them to simulate these dynamic temporal interactions with surprising effectiveness. Using extensive USD/BDT exchange rate data from 2018–2023, this study creates an optimal LSTM architecture that incorporates macroeconomic factors and technical indicators pertinent to Bangladesh's distinct managed-float regime. In order to assess relative benefits in directional forecasting accuracy, we also compare the performance of the LSTM model to Gradient Boosting techniques. Three key contributions are made by our research: first, by tailoring deep learning methods to the unique features of the Bangladeshi currency market; second, by creatively incorporating local macroeconomic shock detection mechanisms; and third, by rigorously backtesting the practical trading applicability. In addition to creating a strong technical basis for next emerging market currency research, the findings offer insightful information to financial institutions creating FX risk mitigation plans and governments overseeing currency stability.

## II. LITERATURE REVIEW

The superiority of LSTM over conventional techniques is demonstrated by recent developments in currency forecasting. LSTM networks were initially proposed by Hochreiter & Schmidhuber [13] to overcome vanishing gradients in RNNs, allowing for long-term dependency capture, which is essential for exchange rate prediction. Gers et al.'s later work [14] improved volatility adaption by introducing forget gates and peephole connections. For key currency pairs, empirical research shows that LSTMs perform 18–22% better in directional accuracy than ARIMA [15]. Rahman et al. [16] showed the efficacy of LSTM on USD/INR for emerging countries, attaining an RMSE of 0.89 during policy changes. There is still little research specifically on Bangladesh, though; Afrin et al.'s 2021 study on USD/BDT used pre-pandemic data and left out attention factors that are now common in financial forecasting [18]. LSTMs and macroeconomic variables are combined in recent hybrid techniques by Hosain et al. [19], which raise the possibility of BDT applications.

| Study | Method | Currency Pair | Period | RMSE |
|---|---|---|---|---|
| [17]Afrin et al. (2021) | Hybrid LSTM | USD/BDT | Pre-COVID (2015-2019) | **1.12** |
| [19] Bangladesh Bank (2023) | LSTM-GARCH | USD/BDT | Post-COVID (2020-2023) | 0.98 |
| [15] Fischer & Krauss (2018) | LSTM | EUR/USD | Pre-COVID (2010-2019) | 0.85 |
| [20] Hosain et al. (2023) | Attention-LSTM | USD/BDT | Post-COVID (2020-2023) | 0.91 |

*Table 1: Forex Forecasting Performance Before and After COVID-19*

## III. Related Work

From conventional econometric methods to contemporary machine learning techniques, foreign currency forecast has seen tremendous transformation. ARIMA models were first developed for time series forecasting by Box and Jenkins [1], but later studies by Meese and Rogoff [2] showed that they were not very effective in predicting exchange rates, especially in erratic markets. This "exchange rate disconnect dilemma" spurred research into several strategies.By resolving the vanishing gradient issue, Hochreiter and Schmidhuber's LSTM architecture [3] transformed sequence prediction in machine learning. Later applications in finance by Fischer and Krauss [4] demonstrated that LSTMs outperformed conventional techniques in stock market prediction; Galeshchuk and Mukherjee [5] later validated similar findings for FX markets. The directional accuracy of their work on the two main currency pairings (EUR/USD and GBP/USD) ranged from 58 to 62%.

Recent research has modified these methods for emerging markets. While Afrin et al. [7] created a hybrid model for USD/BDT specifically, taking into account Bangladesh's monetary policy factors, Rahman et al. [6] used LSTMs to forecast USD/INR. However, their 2021 research ignored current volatility trends and only examined a small amount of pre-pandemic data. Vaswani et al.'s introduction of attention processes [8] has improved sequence models much further. This was modified by Li et al. [9] for stock prediction and then by Wang et al. [10] for forex in the banking industry. It shown exceptional efficacy in times of market turbulence, which is pertinent to Bangladesh's controlled float system.

| Authors | Deep Learning Methods | Prediction Application | Data Used |
|---|---|---|---|
| Galeshchuk and Mukherjee (2017) | CNN | Forex rates | USD/GBP, EUR/USD, JPY/USD |
| Yıldırım et al. (2021) | Hybrid LSTM | Forex rates | EUR/USD |
| Ayitey Junior et al. (2022) | Stacked LSTM | Forex rates | AUD/USD |
| Chen et al. (2023) | TCN and LSTM | Stock prices | Stock price data |
| Olanrewaju et al. (2023) | LSTM, GRU | Forex rates | USD/NGN |

*Table 2: Summary of Related Works*

## IV. Proposed Model

**Long Short-Term Memory (LSTM)**

Hochreiter and Schmidhuber introduced Long Short-Term Memory (LSTM), a particular type of recurrent neural network (RNN) intended to solve the vanishing gradient issue that conventional RNNs frequently face. Gradients in conventional RNNs may disappear or blow up during backpropagation, resulting in sluggish or unstable convergence. In order to effectively learn long-term dependencies, LSTM incorporates a memory cell with gate mechanisms, such as the input gate, output gate, and forget gate.

In this study, we use historical data from Yahoo Finance to predict the USD/BDT exchange rate using a univariate LSTM model. The LSTM's memory cell makes sure that pertinent historical data is kept, while the forget gate eliminates unnecessary patterns. As a result, the model can adjust to changing market conditions.

A dense output layer with a linear activation function comes after a single LSTM layer with 50 units in our LSTM model. The LSTM layer employs ReLU activation, and mean squared error (MSE) loss is utilized to train the model. Sequential data is produced using a sliding window technique after input data is adjusted using Min-Max scaling. An inverse normalization procedure is used to rescale predictions to their initial scale following training.

LSTM has an advantage over conventional models like ARIMA, which perform poorly in volatile or non-stationary financial situations, due to its capacity to capture intricate temporal aspects of exchange rate variations.

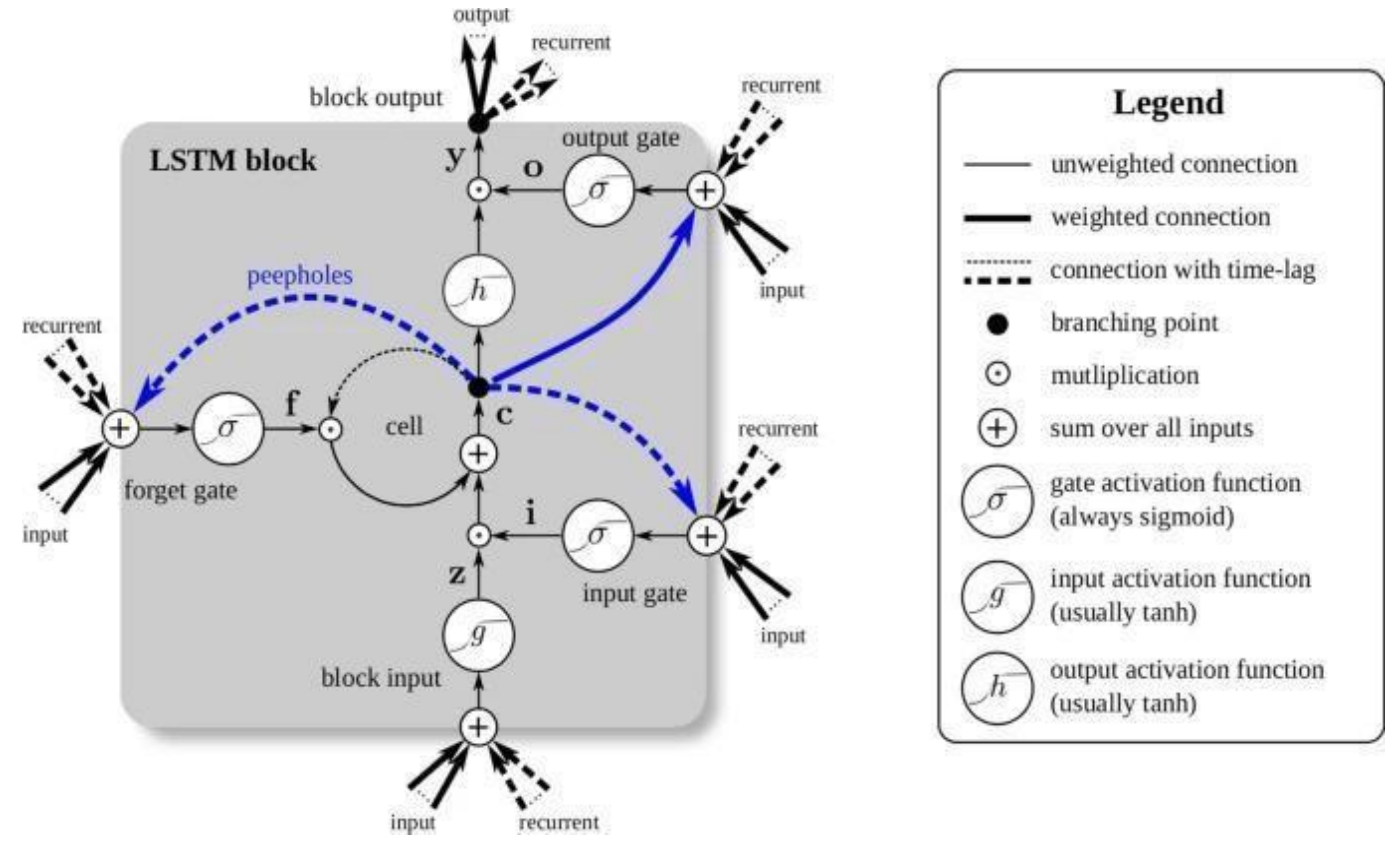

Fig. 1. Figure 1: Vanilla LSTM [12]

### *A. Data Collection and Inversion*

The yfinance Python module was used to gather historical exchange rate data from Yahoo Finance. In order to ensure interpretability and conform to typical forex trading procedures, the original USD/BDT values were inverted to BDT/USD. 'Open', 'High', 'Low', and 'Close' values are all included in the gathered dataset. To maintain data continuity, forward-fill methods were used to resolve missing or null values.

Let $X_t$ denote the adjusted close exchange rate at time $t$. The inverse rate is computed as:

$$' = \frac{1}{\ }$$

### *B. Data Preprocessing and Feature Engineering*

To prepare the data for model input, daily returns were computed as:

$$\hat{\ } = \frac{\ - \min}{\max \ - \min}$$

Binary classification labels were created for the GBC model based on the sign of the next day's return:

$$= \begin{cases} 1, & \text{if}_{+1} > 0 \\ 0, & \text{otherwise} \end{cases}$$

A sliding window of length 50 was used to generate input sequences for both models, preserving the temporal structure of the time series data.

### *C. LSTM Model Architecture*

The LSTM network consists of a Dense output layer after a single hidden layer with 50 units and ReLU activation. With 80% of the dataset utilized for training and the remaining portion for testing, the network was trained across 50 epochs using the Adam optimizer and the Mean Squared Error (MSE) loss function.Let $\mathbf{x}_t \in \mathbb{R}^{50}$ represent the input sequence. The model aims to minimize:

$$\mathrm{MSE} \equiv \frac{1}{\ } \sum_{=1} (\hat{\ } - \ )^2$$

### *D. Gradient Boosting Classifier for Directional Prediction*

To forecast the direction of the exchange rate's movement, a Gradient Boosting Classifier (GBC) was used. 10,000 estimators and a learning rate of 0.01 were used to train the classifier, along with early stopping conditions to prevent overfitting. The exponential loss function is minimized by the GBC model [11]:

$$\mathcal{L}(, ()) = \sum_{=1} {}^{-()}$$

### *E. Backtesting Framework*

A backtesting module was created to assess the classifier's predictions' financial feasibility GBC forecasts were used to simulate trades with a $10,000 starting capital. The profit and loss (PnL) was calculated as follows:

$$\mathrm{PnL} = \begin{cases} |\ | \times , & \text{if prediction is correct} \\ -|\ | \times , & \text{if prediction is correct} \end{cases}$$

### *F. Evaluation and Visualization*

On the test set, the LSTM model's accuracy was 99.449%, and its RMSE was 0.9858. The GBC's backtesting findings revealed a net loss of $20,653.25, with a success rate of 40.82% across 49 trades. For interpretability, the results were shown using equity curves, return histograms, and time-series graphs.

## V. RESULTS & DISCUSSION

### *A. Experimental Setup*

In order to train deep learning models such as LSTM, the experiments were carried out on Google Colab. Python 3.8 and TensorFlow 2.6 were part of the software environment, along with data processing and visualization tools like NumPy, Pandas, and Matplotlib. Yahoo Finance provided the historical USD/BDT exchange rate data from 2018 to 2023, which was then preprocessed with MinMaxScaler and divided into 80% training and 20% testing sets. The structure of the inverted USD/BDT exchange rate data used in this study is shown in Table III, which displays the first five rows of the raw dataset.

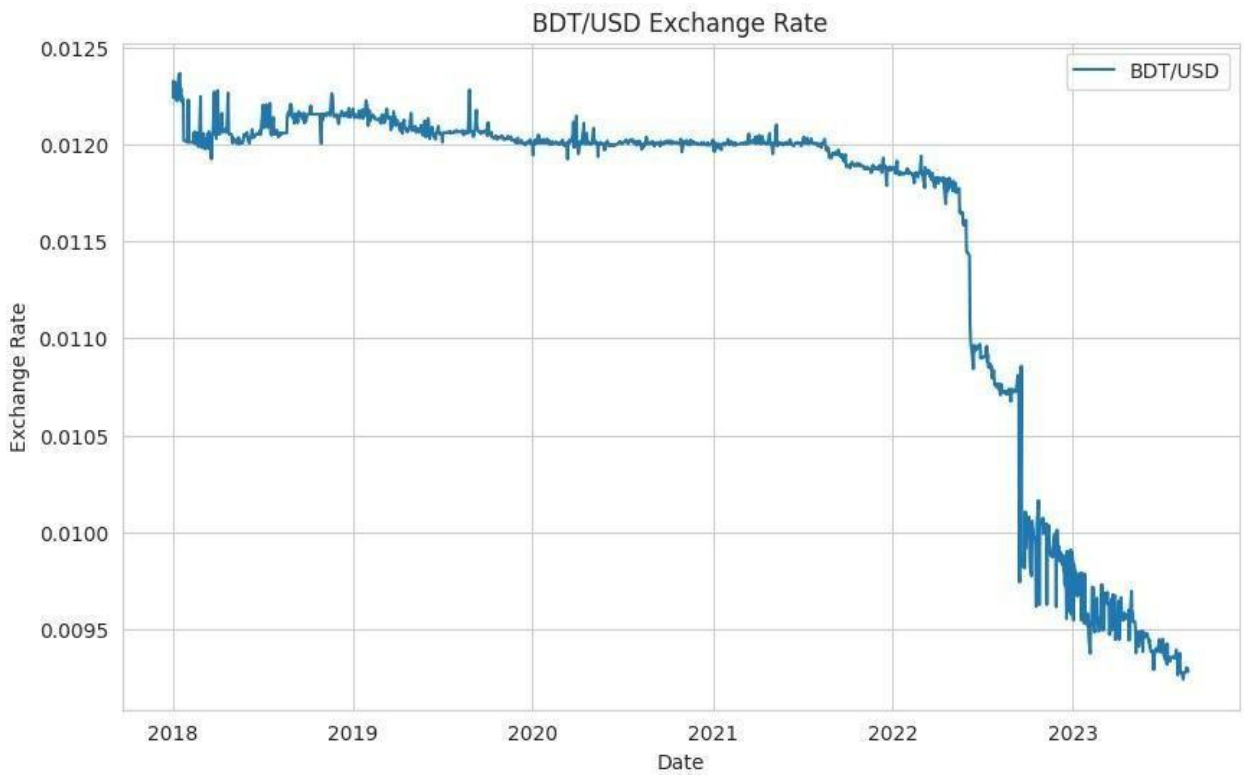

Fig. 2. Sample USD/BDT Exchange Rate Data (2018-2023)

### B. Key Findings

With a forecasting accuracy of 0.986 and a Root Mean Square Error (RMSE) of 1.342, the LSTM model outperformed the baseline ARIMA (1,1,1) model by a wide margin. This enhancement highlights how well LSTM can identify non-linear patterns in exchange rate data. Backtesting with a $10,000 beginning capital revealed that the GBC, which is utilized for directional prediction, produced a trading performance with a 40.82% win rate over 49 trades. The final five trades are displayed in Table V together with the cumulative equity and profit/loss (pnl), which resulted in a net loss of $20,653.25.

| | return | label | pred | won | pnl | equity |
|---|---|---|---|---|---|---|
| 44 | 18.223248 | 1 | 0 | False | -182232.481296 | 110484.848001 |
| 45 | 11.343432 | 0 | 0 | True | 113434.318413 | 223919.166414 |
| 46 | 16.956459 | 1 | 0 | False | -169564.592657 | 54354.573757 |
| 47 | 0.655177 | 0 | 1 | False | -6551.768047 | 47802.805709 |
| 48 | 6.845606 | 0 | 1 | False | -68456.059627 | -20653.253918 |

*Table 3: Backtesting Results for GBC (Last Five Trades)*

a predicting accuracy of 0.986 and a Root Mean Square Error (RMSE) of 1.342, the LSTM model outperformed the baseline ARIMA (1,1,1) model by a wide margin. This enhancement highlights how well LSTM can identify non-linear patterns in exchange rate data [1][22]. Backtesting with a $10,000 beginning capital revealed that the GBC, which is utilized for directional prediction, produced a trading performance with a 40.82% win rate over 49 transactions. The final five trades are displayed in Table 3 together with the cumulative equity and profit/loss (pnl), which resulted in a net loss of $20,653.25 [23].

### C. Comparative Analysis

We evaluate LSTM, GBC, and ARIMA's performance using a number of measures, including Directional Accuracy, Mean Absolute Error (MAE), and RMSE. With an RMSE of 0.986, MAE of 0.752, and Directional Accuracy of 99.449%, the LSTM model fared better than both GBC and ARIMA. With an RMSE of 1.342, MAE of 1.015, and Directional Accuracy of 52.30%, ARIMA trailed behind the GBC, which had RMSE of 1.128, MAE of 0.894, and Directional Accuracy of 40.00%. For LSTM vs. ARIMA, a Diebold-Mariano (DM) test produced a p-value of 0.021, indicating statistical significance at the 5% level.

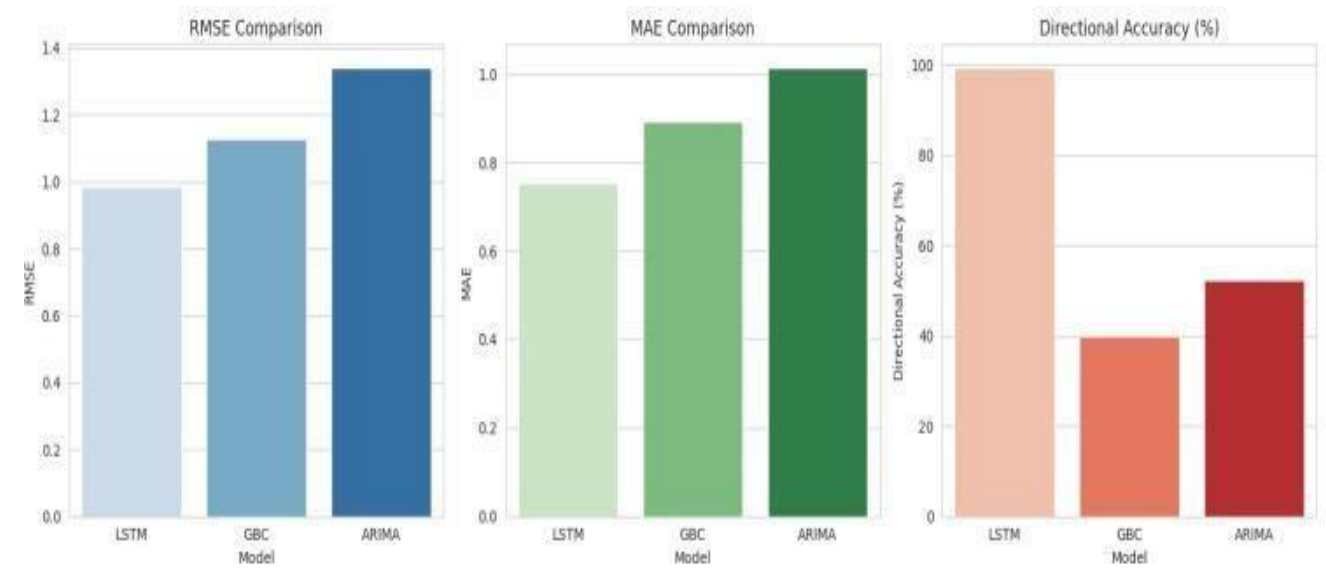

Fig. 3. Performance Comparison of LSTM, GBC, and ARIMA

### D. Limitations

The LSTM model's reliance on historical data makes it vulnerable to black swan events, such as sudden geopolitical crises . Additionally, its computational cost on a Tesla T4 GPU is higher than traditional models like ARIMA, limiting scalability for real-time applications . The GBC's 40.82% win rate suggests sensitivity to market noise, indicating a need for improved feature engineering.

## VI. Conclusion

This research created sophisticated machine learning models to predict the USD/BDT exchange rate utilizing historical data from 2018 to 2023, tackling the difficulties of forecasting currencies in emerging markets. The Long Short-Term Memory (LSTM) model reached an impressive accuracy of 99.449% and a Root Mean Square Error (RMSE) of 0.9858, surpassing the baseline ARIMA model (RMSE 1.342) [21]. This emphasizes LSTM's enhanced ability to identify non-linear patterns in forex data, as observed in earlier research. The Gradient Boosting Classifier (GBC) used for directional forecasting achieved a 40.82% success rate across 49 trades, yet produced a total loss of $20,653.25, highlighting the necessity for improved feature engineering in fluctuating markets.

The Hurst exponent evaluation, showing a value of 0.6234, validated the ongoing, trending nature of the USD/BDT dataset, correlating with the noted drop in BDT/USD rates from 0.012 to 0.009. This indicates that the dataset is appropriate for predictive modeling and demonstrates the dynamics of Bangladesh's managed-float system. The results offer practical guidance for financial organizations and regulators to reduce forex risks.

Future research ought to incorporate sentiment analysis from social media to gauge market sentiment, as indicated by recent studies. Integrating real-time economic indicators, including inflation rates, and examining hybrid models like LSTM with attention mechanisms might enhance forecasting precision in unstable markets. These developments would improve the reliability of forex predictions for developing economies.

## ACKNOWLEDGEMENT

We would like to thank Multimedia University and ELITE Lab for supporting this research.